\documentclass[aps,prl,twocolumn,showpacs,superscriptaddress,nofootinbib,groupedaddress]{revtex4} 
\usepackage{pstricks}
\usepackage{color}
\usepackage{graphicx}
\usepackage{amsmath}
\usepackage{xspace}
\usepackage{amssymb}
\usepackage[colorlinks=true,allcolors=darkred, pdfborder={0 0 0}]{hyperref}
\usepackage[T1]{fontenc}
\usepackage[latin9]{inputenc}
\usepackage{array}
\usepackage{ulem}

\usepackage{booktabs}
\renewcommand{\toprule}{\specialrule{1.5pt}{0em}{0em}}
\renewcommand{\bottomrule}{\specialrule{1.5pt}{0em}{0em}}

\definecolor{darkred}{rgb}{0.6,0,0}

\definecolor{darkred}{rgb}{0.6471, 0.1098, 0.1882} 

\graphicspath{{figs/}}

\def\gsim{\raise0.3ex\hbox{$\;>$\kern-0.75em\raise-1.1ex\hbox{$\sim\;$}}}
\def\lsim{\raise0.3ex\hbox{$\;<$\kern-0.75em\raise-1.1ex\hbox{$\sim\;$}}}

\def\beqn#1{\begin{equation}\label{eq:#1}}
\def\eeqn{\end{equation}}

\def\beqa#1{\begin{eqnarray}\label{eq:#1}}
\def\eeqa{\end{eqnarray}}

\newcommand{\fig}[1]{figure~(\ref{fig:#1})}

\newcommand{\tab}[1]{table~(\ref{tab:#1})}

\newcommand{\eq}[1]{eq.~(\ref{eq:#1})}

\def\TrTrOne{SU(3)_C \otimes SU(3)_L \otimes U(1)_X}
\def\EW{SU(2)_L \otimes U(1)_Y}

\def\z2{\mathbb{Z}_2}
\def\hc{\mathrm{h.c.}}
\def\ie{i.e.,\xspace}

\def\vev#1{\left\langle #1\right\rangle}

\newcommand{\order}[1]{\mathcal{O}(#1)}

\def\nn{\nonumber}

\def\one{\ensuremath{\mathbf{1}}}

\def\three{\ensuremath{\mathbf{3}}}
\def\threeS{\ensuremath{\mathbf{3^*}}}

\newcommand{\gev}{\,\mathrm{GeV}}
\newcommand{\tev}{\,\mathrm{TeV}}

\newcommand{\ev}{\,\mathrm{eV}}

\begin{document}

\title{Predicting charged lepton flavor violation from 3-3-1 gauge symmetry}
\author{Sofiane M. Boucenna}
\email{boucenna@lnf.infn.it}
\affiliation{Instituto de F\'{\i}sica Corpuscular (CSIC-Universitat de Val\`{e}ncia), Apdo. 22085, E-46071 Valencia, Spain.}
\affiliation{INFN, Laboratori Nazionali di Frascati, C.P. 13, 100044 Frascati, Italy.}
\author{Jos\'e W.F. Valle}
\email{valle@ific.uv.es}
\affiliation{Instituto de F\'{\i}sica Corpuscular (CSIC-Universitat de Val\`{e}ncia), Apdo. 22085, E-46071 Valencia, Spain.}
\author{Avelino Vicente}
\email{avelino.vicente@ulg.ac.be}
\affiliation{Instituto de F\'{\i}sica Corpuscular (CSIC-Universitat de Val\`{e}ncia), Apdo. 22085, E-46071 Valencia, Spain.}
\affiliation{IFPA, Dep. AGO, Universit\'e Li\`ege, Bat B5, Sart-Tilman B-4000, Li\`ege 1, Belgium}
\date{\today}

%CHECK
%CHECK

%%%%%%%%%%%%%%%%%%%

\begin{abstract}
\noindent
 
The simplest realization of the inverse seesaw mechanism in a
$\TrTrOne$ gauge theory offers striking flavor correlations between
rare charged lepton flavor violating decays and the measured neutrino
oscillations parameters. The predictions follow from the gauge
structure itself without the need for any flavor symmetry. Such tight
complementarity between charged lepton flavor violation and neutrino
oscillations renders the scenario strictly testable.
\end{abstract}

\maketitle

\section*{Preliminaries}

Beyond the discovery of the Higgs
boson~\cite{Aad:2012tfa,Chatrchyan:2012ufa} no signs of genuine new
physics have shown up so far at high energies.
However, the existence of new physics has been established with the
discovery of neutrino
oscillations~\cite{Maltoni:2004ei,Beringer:1900zz}, implying the
existence of lepton flavor violation and nonzero neutrino masses.
Unraveling the origin of the latter constitutes one of the main
challenges of particle physics.
While the prevailing view is that neutrino masses arise from physics
associated with unification, they might as well signal novel TeV-scale
physics leading to potentially large charged lepton flavor violating
(LFV) rates and possibly also new phenomena testable at the
LHC~\cite{Boucenna:2014zba}.
In this case it could well be that new physics may actually show up
mainly in the form of lepton flavor violation, boosting the motivation
to search for charged LFV phenomena such as the rare decay $\mu \to e
\gamma$.
In fact, the current limit $\text{BR}(\mu \to e \gamma) < 5.7 \times
10^{-13}$~\cite{Adam:2013mnn} already puts severe constraints on
models of new physics.

Models based on the $\TrTrOne$ gauge theory (3-3-1) constitute a
minimal extension of the Standard Model (SM) that accounts for the
existence of three families of fermions, the same as the number of
colors~\cite{Singer:1980sw,valle:1983dk}.
They provide an economical scheme to generate tiny neutrino masses
radiatively from TeV scale physics~\cite{Boucenna:2014ela} and could
lead to successful gauge coupling unification through neutrino masses
and TeV scale physics~\cite{Boucenna:2014dia}. Moreover, they
naturally solve the strong $\mathcal{CP}$ problem by including in an
elegant way the Peccei-Quinn symmetry~\cite{Pal:1994ba,Peccei:1977hh}.

Here we focus on the phenomenology of lepton flavor violation in the
$\TrTrOne$ schemes. 
For definiteness we focus on the simplest implementation of the
inverse seesaw mechanism within the 3-3-1 model. We show that it
offers striking flavor correlations between rare charged lepton flavor
violating decays and the measured neutrino oscillations
parameters. Such predictions result from the gauge theory structure
itself without the need for imposing any specific flavor symmetry. We
analyze the complementarity between charged LFV and neutrino
oscillations, a feature which may render the 3-3-1 scenario strictly
testable within the upcoming generation of LFV searches.

\section*{The Model}

We consider a variant of the model introduced
in~\cite{Boucenna:2014ela} in which neutrinos get masses via the
inverse seesaw mechanism instead of quantum corrections. The model is
based on the $\TrTrOne$ gauge symmetry, extended with a global
$U(1)_{\mathcal{L}}$ so as to consistently define lepton
number~\footnote{For other inverse seesaw constructions within 3-3-1
  scenarios see \cite{Catano:2012kw,Dias:2012xp}.}. We also invoke an
auxiliary parity symmetry in order to ensure a realistic quark mass
spectrum. The model contains three generations of lepton triplets
($\psi_L$), two generations of quark triplets ($Q_L^{1,2}$), one
generation of quark anti-triplet ($Q_L^3$), along, of course, with
their iso-singlet right-handed partners, and accompanied by three
generations of neutral fermion singlets ($S$). The gauge symmetry
breaking is implemented through three scalar anti-triplets
($\phi_{1,2,3}$).  The particle content of the model is summarized
in~\tab{content}.  The fundamental fermions interact through the
exchange of $17$ gauge bosons: the $8$ gluons of $SU(3)_C$, the $8$
``weak'' $W_i$ bosons associated to $SU(3)_L$ ($4$ of which form $2$
electrically charged bosons, and the rest are neutral), and the $B$
boson associated to $U(1)_X$.\\

The lepton representations in \tab{content} can be decomposed as:
\begin{equation} \label{eq:frep}
\psi_L = \left( \begin{array}{c}
\ell^- \\
- \nu \\
N^c \end{array} \right)_L^{e,\mu,\tau}\,,
\end{equation}
{ where we identify $N_L^c \equiv \left( \nu_R \right)^c$~\cite{valle:1983dk}.}

In the scalar sector, on the other hand, we have:
\begin{equation} \label{eq:srep}
\phi_1 = \left( \begin{array}{c}
\phi_1^0 \\
- \phi_1^- \\
\tilde \phi_1^- \end{array} \right) \,, \quad \phi_2 = \left( \begin{array}{c}
\phi_2^+ \\
- \phi_2^0 \\
\tilde \phi_2^0 \end{array} \right) \,, \quad \phi_3 = \left( \begin{array}{c}
\phi_3^+ \\
- \phi_3^0 \\
\tilde \phi_3^0 \end{array} \right) \,.
\end{equation}

\noindent { After electroweak symmetry breaking,} the electric charge
and lepton number assignments of the particles of the model follow
from the action of the operators:
\begin{eqnarray}
\label{eq:QL}
Q&=&T_3+\frac{1}{\sqrt{3}}T_8+X \,; \\
L&=&\frac{4}{\sqrt{3}} T_8+\mathcal{L} \,.
\end{eqnarray}

%%%%%%%%%%%%%%%%%%%%PARTICLE CONTENT%%%%%%%%%%%%%%%%%%%%%%%%%%%%%%%%%%%%%%%%
\begin{table}[!t]
\centering
\begin{tabular}{ c | c c c c c c c c c | c c c}
\toprule
& $\psi_L$ & $l_R$ & $Q_L^{1,2}$ & $Q_L^3$ & $U_R$ & $t^\prime_R$&$D_R$ &$\hat d_R$& $S$  & $\phi_1$ & $\phi_2$ & $\phi_3$ \\ 
\midrule
$SU(3)_C$ & $\one$ & $\one$ & $\three$ & $\three$ & $\three$ & $\three$ &$\three$&$\three$&  $\one$  & $\one$ & $\one$ & $\one$\\
$SU(3)_L$ & $\threeS$ & $\one$ & $\three$ & $\threeS$ & $\one$ & $\one$ & $\one$ & $\one$ & $\one$ & $\threeS$ & $\threeS$ & $\threeS$ \\
$U(1)_X$ & $- \frac{1}{3}$ & $-1$ & $0$ & $+ \frac{1}{3}$ & $+ \frac{2}{3}$ & $+ \frac{2}{3}$ & $- \frac{1}{3}$ &$- \frac{1}{3}$ & $0$ & $+ \frac{2}{3}$ & $- \frac{1}{3}$ & $- \frac{1}{3}$\\[1mm]
\midrule
$U(1)_{\mathcal{L}}$ & $- \frac{1}{3}$ & $-1$ & $- \frac{2}{3}$ & $+ \frac{2}{3}$ & $0$ & $0$ & $0$& $0$ & $+1$ & $+ \frac{2}{3}$ & $- \frac{4}{3}$ & $+ \frac{2}{3}$ \\
% \hline
$\z2$ & $+$ & $+$ & $+$  & $-$ & $+$&$-$ & $-$ & $+$& $+$  & $+$ &$+$&$-$ \\
\bottomrule
%generations & $3$ & $3$ & $2$ & $1$ & $3$ & $1$ & $3$ &$2$& $3$  \\
%\hline
\end{tabular}
\caption{Particle content of the model. Here $U_R\equiv \{u_R, c_R, t_R\}$, $D_R\equiv \{d_R,s_R,b_R\}$ and  $\hat d_R \equiv ( d^\prime_R, s^\prime_R)$.}
\label{tab:content}
\end{table}
%%%%%%%%%%%%%%%%%%%%%%%%%%%%%%%%%%%%%%%%%%%%%%%%%%%%%%%%%%%%%%%%%%%%%%%%%%%%%%

The relevant terms in the Lagrangian for leptons are:
\begin{equation}
\label{eq:laglep}
-\mathcal{L}_{\text{lep}} = y^\ell \bar \psi_L l_R \phi_1 
+ y^a \overline{\psi_L^c} \psi_L \phi_1 
+ y^s \bar \psi_L  S \phi_2 
+ \frac{m_S}{2} \, \overline{S^c}S + \hc \, ,
\end{equation}
where $y^\ell$ and $y^s$ are generic $3 \times 3$ matrices, while
$y^a$ is anti-symmetric and $m_S$ is the $3 \times 3$ Majorana mass
term for the singlets $S$ (symmetric, due to the Pauli
principle). 

\section*{Scalar potential and symmetry breaking}
\label{sec:scalar}

The scalar potential of the model can be written as:
\begin{eqnarray}
\label{eq:V}
V  &=& \sum_i \mu_i^2 |\phi_i|^2 + \lambda_i |\phi_i|^4 + \sum_{i \ne j} \lambda_{ij} |\phi_i|^2|\phi_j|^2 \nonumber\\
&& + f\, (\phi_1 \phi_2 \phi_3+\hc) + m_{\text{s}}^2 \, (\phi_2^\ast \phi_3 +\hc)   \, , 
\end{eqnarray}
where $\mu_{1,2,3}$, $f$ and $m_{\text{s}}$ are parameters with
dimensions of mass. The two latter couplings break the $\z2$
softly. For simplicity we denote all the dimensionless couplings by
$\lambda$ and take $m_{\text{s}}=0$.

In full generality, the scalars of the model are allowed to take
vacuum expectation values (VEVs) in the following directions
$\vev{\phi_1}^T=(k_1,0,0)/\sqrt{2}$,
$\vev{\phi_2}^T=(0,k_3,n)/\sqrt{2}$, and
$\vev{\phi_3}^T=(0,k_2,n^\prime)/\sqrt{2}$. However, in order to
recover the SM as a low energy limit, we assume the hierarchy
$k_{1,2,3} \ll n, n^\prime$.  Moreover, we assume: $k_3 = n^\prime
=0$, which { together with the $\z2$ symmetry} guarantees the
existence of a simple pattern of realistic quark masses (see below).

We define the covariant derivative in the usual way as:
\begin{equation}
D_\mu = \partial_\mu - i \, \sum_{\text{groups}} g A_\mu^a T_a \, ,
\end{equation}
where $A_\mu^a$ is the gauge boson, $T_a$ are the generators of the
group and the sum extends over all gauge groups included in $SU(3)_c
\otimes SU(3)_L \otimes U(1)_X$.
Assuming $k_1 \sim k_2 \equiv k \ll n$, and keeping only the leading
order terms, one finds that the mass spectrum of the charged scalars
is given as: \footnote{We identify the corresponding (approximate) eigenstates between parentheses.}
\begin{eqnarray}
M^2(\phi_2^\pm) &=&0 \,,\\
M^2((\phi_1^\pm + \phi_3^\pm)/\sqrt{2})&=&0 \,,\\
M^2((\phi_1^\pm - \phi_3^\pm)/\sqrt{2})&\sim& \frac{1}{\sqrt{2}}f\, n \,,\\ 
M^2(\tilde \phi_1^\pm) &\sim& \sqrt{2} f\, n \,.%\\
\end{eqnarray}
On the other hand the masses of the neutral CP-even scalars are, up to corrections of $\order{k^2}$:
\begin{eqnarray}
M^2(\Re(\phi_1^0 + \phi_3^0)/\sqrt{2})&\sim&(2 \lambda + \sqrt{2}\, \frac{f}{n} - \frac{1}{2}\,\frac{f^2}{\lambda n^2} ) k^2 \nn\\
\\
M^2(\Re(\phi_1^0 - \phi_3^0)/\sqrt{2})&\sim&  \frac{1}{\sqrt{2}} f \, n\,,\\
M^2(\Re\phi_2^0)&=&0    \,,\\
M^2(\Re\tilde \phi_2) &\sim&2 \lambda\, n^2 \,,\\
M^2(\Re\tilde \phi_3) &\sim& \sqrt{2} f \, n  \,.%\\
\end{eqnarray}
Finally, the masses of the neutral CP-odd scalars at leading order are
given as:
% %
\begin{eqnarray}
% M^2(\phi_2^0)&\sim&0    \,,\\
M^2(\Im(\phi_1^0 + \phi_3^0)/\sqrt{2})&\sim& \sqrt{2} f\,n \,,\\
M^2(\Im(\phi_1^0 - \phi_3^0)/\sqrt{2})&=& 0 \,,\\
M^2(\Im\phi_2^0) &=&0 \,,\\
M^2(\Im\tilde \phi_2) &=&0 \,,\\
M^2(\Im\tilde \phi_3) &\sim& \frac{1}{\sqrt{2}} f \, n  \,.%\\
\end{eqnarray}
The massless scalars found in the above equations correspond to the
degrees of freedom `eaten-up' by the charged and neutral gauge bosons,
respectively, which acquire the following masses:
\begin{eqnarray}
m_{W}^2  &=& \frac{1}{2} \, g_2^2 \, k^2 \,, \\
m_{W'}^2  &=& \frac{1}{4} \, g_2^2 \, n^2 \,,\\
m_Z^2\,\,    &=&  \frac{g_2^2(4 g_1^2+3 g_2^2)}{2 \left(g_1^2+3 g_2^2 \right)}\,k^2  \,, \\
m_{Z'}^2 &=& \frac{1}{9} \, (g_1^2+ 3 g_2^2) \, n^2 \,, \\
m_{X}^2  &=& m_{Y}^2 \, = \, \frac{1}{4} \, g_2^2 \, n^2\,.
\end{eqnarray}

Notice that since $\tilde \phi_2^0$ is singlet under the $SU(2)_L$
subgroup contained in $SU(3)_L$, the VEV $n$ { will control the four
  new gauge bosons masses} and break $SU(3)_L$ to $SU(2)_L$.  On the
other hand, $\EW$ is broken at the electroweak scale by the $k_1$ and
$k_2$ VEVs down to the electromagnetic $U(1)_Q$ symmetry.
For $f\sim n$ all the scalars of the model are naturally
heavy, except one state that we can identify with the SM Higgs boson,
\ie $H\equiv (\phi_1^0 + \phi_3^0)/\sqrt{2}$, in good
approximation. Indeed, its couplings to the fermions confirm that the
state $H$ is the one that gives mass to SM fermions.

\section*{Quark sector}
\label{sec:quark}

We now turn to the quark sector. From the symmetries of the model, see \tab{content}, it follows that the quark Lagrangian is given by:
\begin{eqnarray}
\mathcal{L}_{\text{quarks}} &=& \bar Q_L^{1,2}\, y^u  U_R \phi_1^\ast +
\bar Q_L^{1,2}\, y^d D_R \phi_3^\ast + \bar Q_L^{1,2}\, \bar y^d   \hat d_R \phi_2^\ast \nn\\
&&+\,\bar Q_L^3\, \tilde y^u  U_R \phi_3 +
\bar Q_L^{3}\, \tilde y^d D_R \phi_1+
 \bar Q_L^3\, \bar y^u  t^\prime_R \phi_2 \nonumber    \nonumber\\
&&+\, \hc\,,
\label{eq:lagqrk}
\end{eqnarray}
where we defined $\hat d_R \equiv ( d^\prime_R, s^\prime_R)$, $U_R\equiv \{u_R, c_R, t_R\}$ and $D_R\equiv \{d_R,s_R,b_R\}$.
This Lagrangian leads to the following mass matrices:
\begin{equation}
M_d = - \frac{1}{\sqrt{2}} \left( \begin{array}{ccccc}
y^d_{11} \, k_2 & y^d_{12} \, k_2 & y^d_{13} \, k_2 & 0 & 0 \\
y^d_{21} \, k_2 & y^d_{22} \, k_2 & y^d_{23} \, k_2 & 0 & 0 \\
\tilde y^d_{11} \, k_1 & \tilde y^d_{12} \, k_1 & \tilde y^d_{13} \, k_1 & 0 & 0 \\
0 & 0 & 0 & \bar y^d_{14} \, n & \bar y^d_{15} \, n \\
0 & 0 & 0 & \bar y^d_{24} \, n & \bar y^d_{25} \, n
\end{array} \right)\,,
\end{equation}
\begin{equation}
M_u = - \frac{1}{\sqrt{2}} \left( \begin{array}{cccc}
y^u_{11} \, k_1 & y^u_{12} \, k_1 & y^u_{13} \, k_1 & 0 \\
y^u_{21} \, k_1 & y^u_{22} \, k_1 & y^u_{23} \, k_1 & 0 \\
\tilde y^u_{11} \, k_2 & \tilde y^u_{12} \, k_2 & \tilde y^u_{13} \, k_2 & 0 \\
0 & 0 & 0 & \bar y^u_{14} \, n
\end{array} \right) \, .
\end{equation}
Thanks to the $\z2$ symmetry, the SM and exotic sub-sectors are
independent of each other and can be adjusted individually to easily
obtain a realistic quark sector and heavy exotic quarks at the same
time.

\section*{Neutrino masses and inverse seesaw mechanism}
\label{sec:numass} 

{ The presence of the small term $\overline{S^c} S$, in \eq{laglep},
  explicitly breaks $U(1)_\mathcal{L}$ and provides the seed for
  lepton number violation leading to neutrino masses via the inverse
  seesaw mechanism.}
Indeed, after spontaneous symmetry breaking of the electroweak gauge
group, we get the following $9 \times 9$ neutrino mass matrix, in the
basis ($\nu, N, S$)~\cite{Mohapatra:1986bd}:
\beqn{mnu1}
\mathcal M = %\simeq
\left(
\begin{array}{ccc}
0 & m_D & 0\\
& 0 & M\\
& & m_S
\end{array}
\right)\, , 
\eeqn
where $m_D \equiv \sqrt{2} k_1 \, y^a$ and $M\equiv \frac{1}{\sqrt{2}}
n \, y^s$~\footnote{Note that the matrix in \eq{mnu1} does not depend
  on the conditions imposed on the VEVs. Indeed, even if $k_3 \neq 0$
  the resulting linear seesaw
  term~\cite{Akhmedov:1995ip,Akhmedov:1995vm,Malinsky:2005bi} would
  give only a subleading contribution $\sim m_\nu (M_W/n)^2$}.  The
inverse seesaw-induced light neutrino masses can be written
as~\cite{Mohapatra:1986bd}:
\beqn{inv1}
m_\nu = m_D \, \left( M^T \right)^{-1} \, m_S \, M^{-1} \, m_D^T \, .
\eeqn
Here, the matrix $M$ can be taken diagonal without loss of generality.
Using this freedom and taking into account that $m_D$ is
anti-symmetric, \eq{inv1} can be expressed in terms of an effective
symmetric $3 \times 3$ matrix, ${\widetilde M}^{-1} \equiv M^{-1} \,
m_S \, M^{-1}$, as:
\beqn{inv2}
m_\nu = - m_D \, M^{-1} \, m_S \, M^{-1} \, m_D \equiv - m_D \, {\widetilde M}^{-1} \, m_D \,.
\eeqn
A simple implication of the antisymmetry of the ``Dirac'' entry $m_D$
is that $\text{Det}(m_\nu) = 0$, so that the lightest neutrino in this
model must be massless at the tree level.

\section*{Lepton flavor violation predictions}

Let us now proceed to a simple parameter counting. On the left-hand
side of \eq{inv1} one has $5$ independent complex parameters, since
$\text{Det}(m_\nu) = 0$. In contrast, on the right-hand side
of~\eq{inv1} one has $9$ independent complex parameters: $3$ in $m_D$,
and $6$ elements in ${\widetilde M}$. Therefore, we have $4$ (complex)
relations among the parameters ($y^s, y^a$, and $m_S$).
One can choose as free parameters the $3$ off-diagonal entries of
${\widetilde M}^{-1}$, together with a global scaling factor $\tilde
m$ defined through:
\beqn{eq:mD}
m_D^{ij} = \tilde m^{-1} \, \left( m_\nu^{1j} m_\nu^{2i} - m_\nu^{1i} m_\nu^{2j} \right) \,.
\eeqn
From~\eq{inv1}, we can see that $\tilde m$ scales as $\sqrt{m_\nu^3
  m_S /M^2}$, so that for $m_S \approx 10 \ev$, $M \approx 1 \tev$,
and neutrino masses of $\order{0.1} \ev$, we obtain $\tilde m \approx
10^{-22} \gev$. In contrast, the diagonal entries of ${\widetilde
  M}^{-1}$ are functions of its off-diagonal elements and $m_D$. We
emphasize that \eq{eq:mD} is not an {\it ansatz}, but the most general
solution of \eq{inv2}.

Such a parameterization makes explicit the direct relation between
charged LFV observables and neutrino oscillation parameters, which is
a characteristic feature of our model. Indeed, LFV in this model
arises from the term:
\begin{equation}
\label{eq:lfv}
- \mathcal{L}_{\text{LFV}}= y^a\, \psi_L^T C^{-1}  \psi_L \phi_1 + \hc \,,
\end{equation}
which depends solely upon the coupling $y^a$, hence $m_D$. Using
\eq{eq:mD} together with $m_\nu = U_\nu^\ast \, m_\nu^{\text{diag}} \,
U_\nu^\dagger$, where $U_\nu$ is the leptonic mixing matrix in its
standard parameterization in terms of three mixing angles and the
Dirac phase ($\delta$), one obtains the relevant coupling for
LFV as:
%%%
\beqa{YA}
y^a &=& \left( \begin{array}{ccc}
0 & \tilde y^a_{12} & \tilde y^a_{13} \\
- \tilde y^a_{12} & 0 & \tilde y^a_{23} \\
- \tilde y^a_{13} & - \tilde y^a_{23} & 0
\end{array} \right) \, \times \frac{\sqrt{\Delta m_{\text{atm}}^2}}{\sqrt{2} k_1 \, \tilde m} \nn \\
&& \nn \\
&& \times
\left\{ \begin{array}{c}
\sqrt{\Delta m_{\text{sol}}^2} \qquad\qquad\qquad \text{    (NH)} \\
\phantom{\Delta}\\
\sqrt{\Delta m_{\text{atm}}^2 + \Delta m_{\text{sol}}^2} \qquad \text{(IH)}
\end{array} \right. \,,
\eeqa
for normal (NH) and inverse hierarchies (IH). Here the parameters
$\tilde y^a_{ij}$ are functions of the lepton mixing matrix
parameters, i.e. $\tilde y^a_{ij} = \tilde
y^a_{ij}(\theta_{12},\theta_{13},\theta_{23},\delta)$, and are  given in the Appendix.
% 
% Since the
% resulting expressions turn out to be quite lengthy, we give them in
% the Appendix.
% 
It is remarkable that the Yukawas $y^a$ relevant for
determination of LFV rates are, up to a global scaling factor, fully
determined by the parameters measured in neutrino oscillation
experiments. This allows us to make definite predictions for LFV
observables that can be used to provide an unambiguous test of the
model, as we show below.

\begin{figure}[t!]
\centering \includegraphics[scale=0.28]{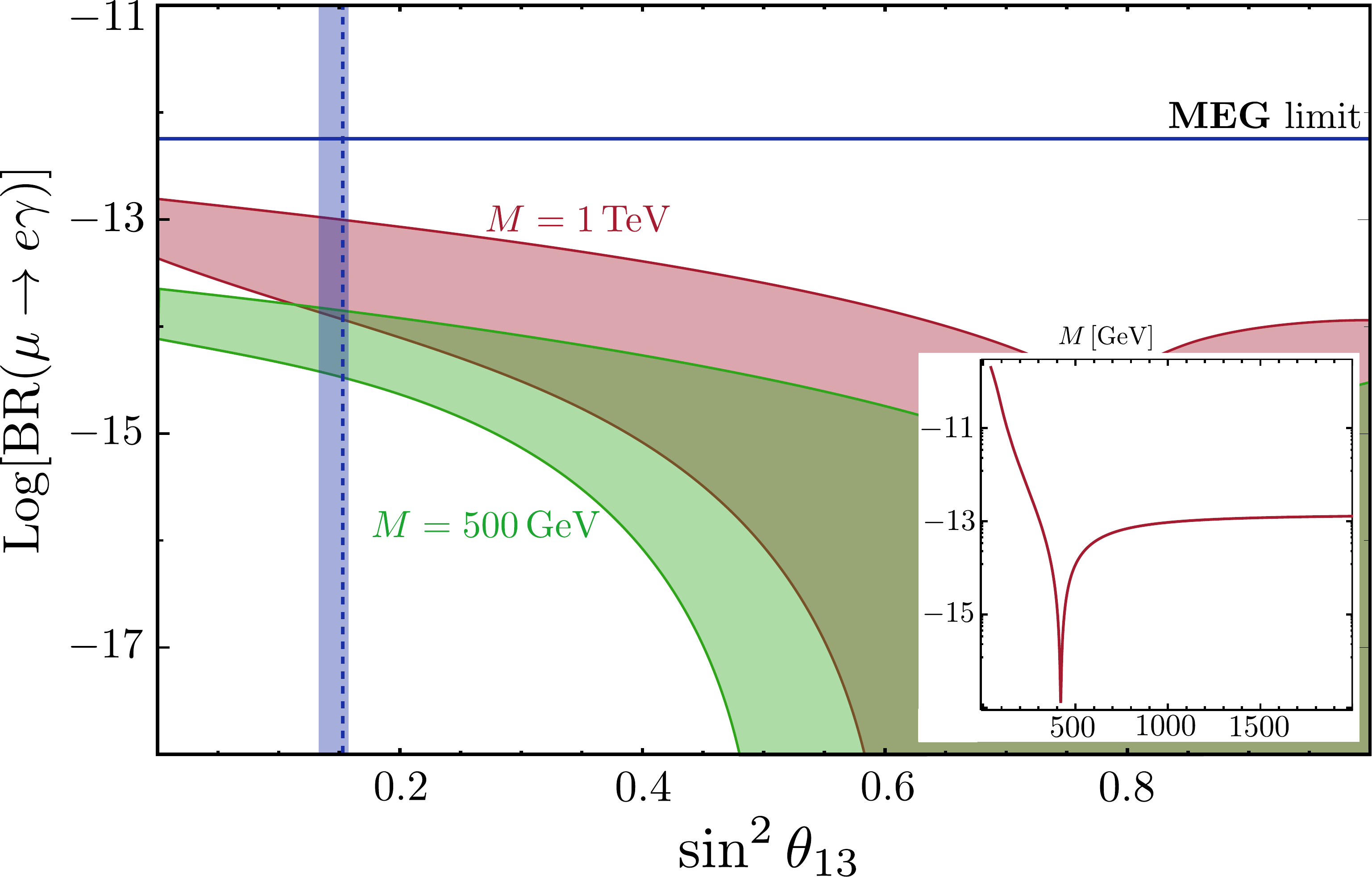}
\caption{ The branching ratio of the decay $\mu \to e \gamma$ versus
  $\sin \theta_{13}$ for $M=500 \gev$ and $1 \tev$. We take
  $\tilde{m}=2 \times 10^{-23}\gev$, $f=2 \tev$, and $n=3 \tev$. The
  vertical band is the $3\sigma$ range reported
  in~\cite{Forero:2014bxa}. The other mixing angles are taken within
  their $3 \sigma $ range~\cite{Forero:2014bxa}. We also show in the
  lower-right corner the $\mu \to e \gamma$ branching ratio as a
  function of $M$, using best-fit values for the mixing angles, as
  given in \cite{Forero:2014bxa}.}
\label{fig:muVS13}
\end{figure}

\subsection*{Radiative $\ell_i \to \ell_j \gamma$ decays}

In order to show the predictive power of the model here we focus, for
definiteness, on flavour-changing leptonic (radiative) decays. This
probe constitutes one of the most important tests of new physics and
has been actively sought after in many experiments.  The branching
ratio (BR) of the decay of the charged lepton $\ell_i \to \ell_j
\gamma$ is given as:
\beqn{lfv}
\text{BR}(\ell_i \to \ell_j \gamma)
=
\frac{m_{\ell_i}^5 \,|\left(y^a \, F \, y^a\right)_{ij}|^2}{\Gamma_{\ell_i}}\,,
\eeqn
where $\Gamma_{\ell_k}$ is the total decay width of $\ell_k$, and $F$
{ is a function that depends on the masses and mixings of all the
  particles running inside the loop (summation over the different
  contributions is implicit here).  We have three different classes of
  contributions: $i$) loops mediated by the new heavy gauge
  bosons. These are suppressed due to the large scale of the breaking
  of $SU(3)_L$ compared to $M_W$; $ii$) contributions from the
  exchange of a charged scalar whose mass is $\sim
  \sqrt{f\,n/\sqrt{2}}$; and finally $iii$) the ``standard'' loop,
  mediated by the SM $W$ boson and neutrinos.
The latter two contributions dominate the $\ell_i \to \ell_j \gamma$ amplitude, with relative sizes depending on the ratio  $f/M$.}\\

This branching ratio depends on the neutrino mixing parameters, the
global scaling factor $\tilde m$, and on the neutrino mass
hierarchy. As can be seen in~\eq{YA}, the off-diagonal entries in
$y^a$ are larger in the case of IH by a
factor $\sim \sqrt{\Delta m_{\text{atm}}^2/\Delta m_{\text{sol}}^2}$ with
respect to NH so that, for the same input
parameters, one has larger LFV effects in IH.\\

{We compute the various relevant LFV observables using {\tt
    FlavorKit}~\cite{Porod:2014xia}~\footnote{This is a computer tool
    based on {\tt SARAH}~\cite{Staub:2013tta} and {\tt
      SPheno}~\cite{Porod:2011nf}, that increases their capability to
    handle flavor observables.}.
The branching ratio of the decay $\mu \to e \gamma$ is shown in
\fig{muVS13} as a function of $\sin\theta_{13}$ for two different
values of the (quasi-Dirac~\cite{valle:1983dk}) right-handed neutrino
mass $M=500 \gev$ and $1 \tev$. The vertical band is the $3\sigma$
range reported in~\cite{Forero:2014bxa}, whereas the other mixing
angles are randomly taken within their $3 \sigma $
range~\cite{Forero:2014bxa}. This figure has been obtained by varying $M$ (by taking different values for the $y^s$
Yukawa couplings) for the fixed parameters $n = 3 \tev$, $f = 2 \tev$, and $\tilde{m}=2 \times 10^{-23}\gev$. We also consider
degenerate right-handed neutrinos, normal hierarchy for the light
neutrinos and a vanishing Dirac CP violating phase. One notices that
the branching ratio is lower for the $M=500 \gev$ case. This is caused
by a partial cancellation between the standard loop, mediated by the
$W$ boson, and the contribution induced by the exchange of charged
scalars. This cancellation takes place for $M \simeq 400 \gev$ and is
explicitly illustrated in the lower-right corner of \fig{muVS13},
where the $\mu \to e \gamma$ branching ratio is shown as a function of
$M$, using best-fit values for the mixing
angles~\cite{Forero:2014bxa}. The main message from \fig{muVS13} is
that $\mu \to e \gamma$ may take place with sizeable rates, close to
the current limit, or even larger. Given the expected sensitivities of
upcoming experiments one finds that the detection of this and other
muon number violating processes might become feasible.

%tau decays
\begin{figure}[t!]
\centering
\includegraphics[scale=0.3]{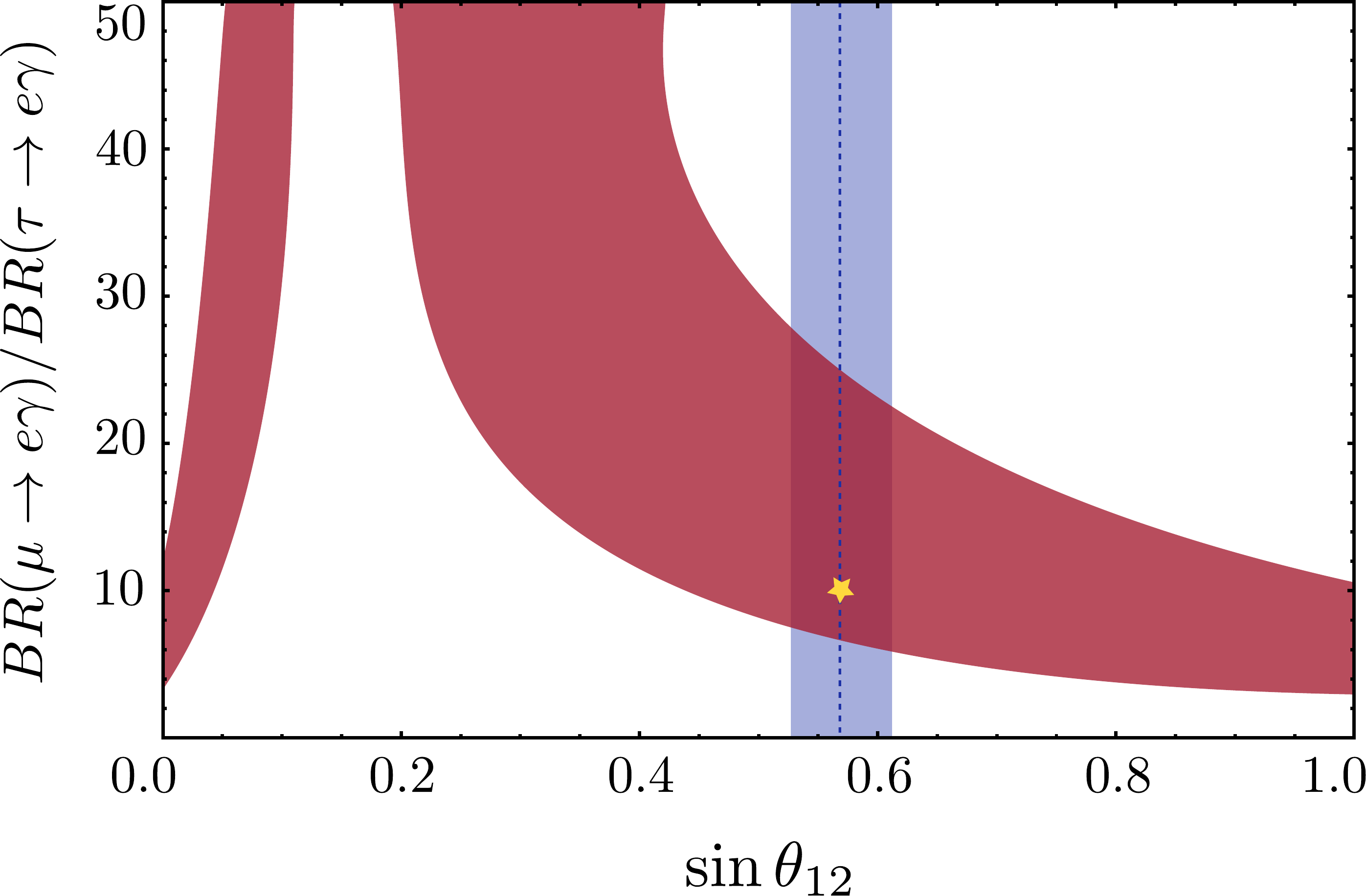} 
\caption{ Ratio of the BR of $\mu$ and $\tau$ decays to $e \gamma$,
  \ie $\text{BR}(\mu \to e \gamma)/\text{BR}(\tau \to e \gamma)$
  versus $\sin \theta_{12}$. The vertical band is the $3\sigma$ range
  given in~\cite{Forero:2014bxa}. The tilted band is obtained by
  varying the other mixing angles within their $3 \sigma$
  range~\cite{Forero:2014bxa}.}
\label{fig:RBRVS12}
\end{figure}

Since the BR depends on a global multiplicative factor, it is
interesting to consider the ratio of branching ratios of LFV
lepton decays.  It follows from~\eq{lfv} that:
\beqn{ratioBRs}
\frac{\text{BR}(\ell_i \to \ell_j \gamma)}
{\text{BR}(\ell_k \to \ell_n \gamma)}
=
\frac{m_{\ell_i}^5}{m_{\ell_k}^5}
\frac{|\left(y^a \, F \, y^a\right)_{ij}|^2/\Gamma_{\ell_i}}
{|\left(y^a \, F \, y^a\right)_{kn}|^2/\Gamma_{\ell_k}} \,.
\eeqn
For the simplest case of nearly degenerate right-handed neutrinos, the
$F$ functions are all equal and cancel out in the fraction. In this case,
\eq{ratioBRs} depends exclusively on the ratios of $|y^a y^a|$, \ie
only on the neutrino mixing angles.
The main advantage of considering the ratio of branching ratios and a
quasi degenerate spectrum is that this leads to clean predictions
which do not depend on the neutrino mass hierarchy nor the loop
functions. 
Indeed, in this simplified scenario, by combining~\eq{YA}
with~\eq{ratioBRs} we obtain the following predictions:
\begin{eqnarray}
\frac{\text{BR}(\mu \to e \gamma)}{\text{BR}(\tau \to e \gamma)} &=& \frac{m_{\mu}^5 \Gamma_{\tau}}{m_{\tau}^5 \Gamma_{\mu}} \frac{|\tilde y^a_{23} \tilde y^a_{13}|^2}{|\tilde y^a_{12} \tilde y^a_{23}|^2}\approx 10 \, , \label{eq:BRratio1} \\
\frac{\text{BR}(\tau \to e \gamma)}{\text{BR}(\tau \to \mu \gamma)} &=& \frac{|\tilde y^a_{12} \tilde y^a_{23}|^2}{|\tilde y^a_{12} \tilde y^a_{13}|^2} \approx 3\,,
\label{eq:BRratio2}
\end{eqnarray}
where we have used the best-fit values for the neutrino parameters as
derived in the global fit of neutrino oscillations given
in~\cite{Forero:2014bxa} and set $\delta = 0$.
So, when the right-handed neutrino spectrum is degenerate, the model
predicts $\text{BR}(\mu \to e \gamma) \gg \text{BR}(\tau \to \ell_i
\gamma)$. Therefore, given the expected sensitivities for $\tau$ LFV
decays~\footnote{The expected Belle II sensitivities for $\tau$
  radiative decays are around $10^{-9}$~\cite{Bevan:2014iga}, whereas
  the current MEG bound on $\text{BR}(\mu \to e \gamma)$ is many
  orders of magnitude stronger, $\text{BR} < 5.7 \times 10^{-13}$.},
the simple observation of $\tau \to \ell_i \gamma$ in one (or several)
of the near future experiments would rule out our simplest degenerate
right-handed neutrino hypothesis. The viable alternative scenario in
such cases would be a hierarchical right-handed neutrinos spectrum,
impliying a non-vanishing contribution of the $F$ loop functions in
the ratio of BRs. In this case the $F$ functions for different flavor
transitions can take very different values, and thus the ratios in
Eq.~\eqref{eq:BRratio2} can clearly depart from their predictions in
the degenerate scenario.\\

As an illustration, in \fig{RBRVS12} we show the ratio of the BR of
$\tau$ and $\mu$ decays to $e \gamma$, namely $\text{BR}(\mu \to e
\gamma)/\text{BR}(\tau \to e \gamma)$ as a function of the solar mixing
parameter $\sin\theta_{12}$.  The other oscillation parameters are
varied randomly within their $3 \sigma$ ranges~\cite{Forero:2014bxa}.
Similarly, the ratio of the BR of leptonic $\tau$ decays, \ie
$\text{BR}(\tau \to \mu \gamma)/\text{BR}(\tau \to e \gamma)$ depends
mainly on the solar mixing parameter.\\

For other LFV processes such as $\ell_i \to 3 \, \ell_j$ and $\mu-e$
conversion in nuclei, our results are qualitatively similar to the
ones found in standard low-scale seesaw
models~\cite{Abada:2014kba}. Loops including neutrinos give the most
important contributions, leading to LFV rates comparable to the ones
for the radiative decay $\ell_i \to \ell_j \gamma$. This will be of
special relevance due to the expected sensitivities in the coming
experiments~\cite{Mihara:2013zna}. The complete study of all LFV
processes is, however, beyond the scope of this paper.

\section*{Conclusions and discussion}

In summary, we have shown how a simple extended $\TrTrOne$ electroweak
gauge symmetry implementing the inverse seesaw mechanism implies
striking flavor correlations between rare charged lepton flavor
violating decays and the measured neutrino oscillations
parameters. The predictions follow simply from the enlarged gauge
structure without any imposed flavor symmetry. Such tight
complementarity between charged LFV and neutrino oscillations renders
the scenario strictly testable. A more detailed study of other LFV
processes will be taken up elsewhere. The scheme also has a
non-trivial structure in the quarks sector since, thanks to the
anomaly cancelation requirements, the Glashow-Iliopoulos-Maiani
mechanism breaks down, leading to a plethora of flavor-changing
neutral currents in the quark
sector~\cite{Buras:2014yna,Buras:2012dp}. Last but not least, the
model presents a rich structure of new physics at the TeV scale that
could be potentially studied in the coming run of the LHC.

\section*{Acknowledgments}

Work supported by the Spanish grants FPA2014-58183-P and Multidark
CSD2009-00064 (MINECO), and the grant PROMETEOII/2014/084 from
Generalitat Valenciana. SMB acknowledges financial support from the
research grant ``Theoretical Astroparticle Physics'' number 2012CPPYP7
under the program PRIN 2012 funded by the Italian ``Ministero
dell'Istruzione, Universit\'a e della Ricerca'' (MIUR) and from the
INFN ``Iniziativa Specifica'' Theoretical Astroparticle Physics
(TAsP-LNF).  We are grateful to F. Staub and W. Porod for valuable
help with {\tt SARAH} and {\tt SPheno}. AV is grateful to
I. Cordero-Carri\'on for valuable discussions and assistance with
matrix gymnastics.

\appendix

\section*{Appendix: $y^a$ Yukawa couplings}
\label{app:yukawa}

We present in this appendix the expressions for the $y^a$
Yukawa couplings. Using the definitions of the $\tilde y^a$ elements in
\eq{YA}, we find
\begin{widetext}
\begin{eqnarray}
\tilde y^a_{12} &=& -\left( e^{i \delta} \cos \theta_{12} \sin \theta_{13} \cos \theta_{23} -\sin \theta_{12} \sin \theta_{23} \right)^2 \, , \\
\tilde y^a_{13} &=& e^{i \delta} \sin \theta_{12} \cos \theta_{12} \sin \theta_{13} \cos \left( 2 \theta_{23} \right) - \sin \theta_{23} \cos \theta_{23} \left( \sin^2 \theta_{12} - e^{2 i \delta} \cos^2 \theta_{12} \sin^2 \theta_{13} \right) \, , \\
\tilde y^a_{23} &=& \cos \theta_{12} \cos \theta_{13} \left( e^{i \delta} \cos \theta_{12} \sin \theta_{13} \cos \theta_{23} -\sin \theta_{12} \sin \theta_{23} \right) \, .
\end{eqnarray}
\end{widetext}

\bibliographystyle{utphys}

\providecommand{\href}[2]{#2}\begingroup\raggedright\endgroup

%\bibliography{refs,newrefs,merged}

\end{document}